\documentclass[twocolumn,runningheads,natbib]{svjour2}

\smartqed  
\usepackage{graphicx}

\newcommand{\papI}{Paper~I}    

\newcommand{\pd}{\partial}           

\newcommand{\Uvec}[1]%
{\ensuremath{\mathbf{e}_{#1}}}  

\newcommand{\GJ}[1]{\ensuremath{#1_\textrm{\tiny GJ}}}    
 
\newcommand{\QPC}{\ensuremath{\theta_\mathrm{pc}}}

\newcommand{\PC}[1]{\ensuremath{#1_\mathrm{pc}}}

\newcommand{\OmF}{\ensuremath{\Omega_\mathrm{F}}}
       
\newcommand{\PsiL}{\ensuremath{\psi_\mathrm{last}}} 

\newcommand{\XL}{\ensuremath{x_0}}
\newcommand{\RLC}{\ensuremath{R_\mathrm{LC}}}     

\graphicspath{{./figures/}}

\journalname{Astrophysics and Space Science}
\begin{document}

\title{Force-free magnetosphere of an aligned rotator
with differential rotation of open magnetic field lines
}

\titlerunning{Force-free magnetosphere of an aligned rotator}   

\author{A.~N.~Timokhin}


\institute{A.~N.~Timokhin\at
              Sternberg Astronomical Institute \\
              Universitetskij pr. 13 \\
              119992 Moscow, Russia
              \email{atim@sai.msu.ru}           
}

\date{Received: date / Accepted: date}

\maketitle

\begin{abstract}
  Here we briefly report on results of self-consis\-tent numerical
  modeling of a differentially rotating force-free magnetosphere of an
  aligned rotator. We show that differential rotation of the open
  field line zone is significant for adjusting of the global structure
  of the magnetosphere to the current density flowing through the
  polar cap cascades. We argue that for most pulsars stationary
  cascades in the polar cap can not support stationary force-free
  configurations of the magnetosphere.
\end{abstract}

\section{Introduction}
\label{intro}

Aligned rotator with the force-free magnetosphere is being considered
as a first order approximation for the real pulsar magnetosphere since
introduction of the model by \citet{GJ}.  Any pulsar model should be
tested for this simplest case. Recently substantial progress in
modeling of pulsar magnetospheres was achieved.  The magnetosphere of
an aligned rotator was modeled using stationary
\citep{CKF,Goodwin/04,Gruzinov:PSR,Timokhin2006:MNRAS1} and
time-dependent
\citep{Komissarov06,Bucciantini06,McKinney:NS:06,Spitkovsky:incl:06}
codes.  The structure of the magnetosphere has been obtained even for
an inclined rotator \citep{Spitkovsky:incl:06}.  In these works the
angular velocity of plasma rotation was assumed to be constant,
although the case when the open field lines rotate with a constant
angular velocity different from the angular velocity of the Neutron
Star (NS) was addressed in some works \citep[see
e.g.][]{Contopoulos05,Beskin/Malyshkin98}.  It was also implicitly
assumed that the current density in the magnetosphere could adjust to
any distribution required by the global structure of the
magnetosphere.  However, in the pulsar magnetosphere current carriers
are produced mainly by the electron-positron cascades and therefore
not every current density distribution can be realized
\citep[see][hereafter \papI]{Timokhin2006:MNRAS1}.  Moreover, in
\papI{} it was shown that the current density distribution supporting
the magnetospheric configuration frequently used in theoretical pulsar
models, that assume the closed field line zone extending up to the
Light Cylinder (LC) and the open field lines rotating with the same
angular velocity as the NS, could not be realized regardless of a
particular model of the polar cap cascades.  Such models require
presence of an electric current flowing against the accelerating
electric field in some parts of the pulsar polar cap, which cannot be
naturally explained.

For the natural stationary configuration of the magnetosphere, when
the last closed field line lies in the equatorial plane and magnetic
field lines become radial at a large distance from the LC (the
so-called Y-configuration), the system has two physical ``degrees of
freedom'', namely i) the size of the closed field line zone and ii)
the angular velocity of rotation of the open field lines.  In \papI{}
the angular velocity of the magnetic field lines was fixed, but the
size of the closed field line zone was varied.  In any of the obtained
configurations the current density is much less than the
Goldreich-Julian (GJ) current density along field lines passing close
to the boundaries of the polar cap (see Fig.~5 in \papI).  However,
the current density, which could be produced by stationary polar cap
cascades operating in space charge limited flow (SCLF) regime, can not
be significantly less than the corresponding GJ current density
\citep[e.g.][]{Harding/Muslimov98}.  On the other hand, the current
density flowing in the polar cap of pulsar having partially screened
polar gap (PSG) cannot be close to zero
\citep{Gil/Melikidze/Geppert:2003}.  Hence, aligned pulsar with
stationary polar cap cascades operating in SCLF or PSG regime cannot
have stationary force-free magnetosphere rotating with a constant
velocity.  As stationary SCLF and PSG are being considered as the most
probable regimes for the polar cap of pulsar, it is worth to find
configuration of the force-free magnetosphere compatible with them.

Here for the first time we consider self-consistently the case of a
differentially rotating force-free magnetosphere of an aligned rotator
in Y-configuration, i.e. we are able to explore \emph{all} possible
configurations of such magnetosphere. Our aim is to study the impact
of differential rotation on the current density distribution in the
force-free magnetosphere of an aligned rotator. We also try to find
combinations of the angular velocity distribution and the size of the
closed field line zone, which could be compatible with stationary
polar cap cascade models.

\section{Equations for differentially rotating force-free magnetosphere}
\label{equation}

Magnetic field lines in the closed zone are equipotential and plasma
there corotates with the NS. In the open field line domain the angular
velocity of plasma $\OmF$ is different from the angular velocity of
the NS rotation $\Omega$. The angular velocity of the open magnetic
field lines $\OmF$ is determined by the potential difference in the
polar cap of pulsar. In stationary cascade models $\OmF$ is close to
$\Omega$ for young pulsars (see Sec.~2.2 in \papI.)

The force-free magnetosphere of an aligned rotator with differential
rotation can be described by a solution of the
so-called pulsar equation, derived by \citet{Okamoto74}. In notations
of \papI{} (see eq.~20 there) it has the form
\begin{eqnarray}
  \label{eq:PsrEq}
  (\beta^2 x^2-1)(\pd_{xx}\psi + \pd_{zz}\psi) +
  \frac{\beta^2 x^2+1}{x} \pd_x \psi -
  && \nonumber\\
  \quad\quad\quad
  - S \frac{d S}{d \psi}
  + x^2 \beta \frac{d \beta}{d \psi} \left( \nabla \psi \right)^2
  & = & 0
  \,.
\end{eqnarray}
Here $\beta \equiv \OmF/\Omega$, $S$ and $\psi$ are normalized
poloidal current and magnetic flux functions correspondingly. $z$ axis
is co-aligned with $\vec{\Omega}$. All coordinates are normalized to
the LC radius of a corotating magnetosphere $\RLC\equiv c/\Omega$. The
last term in the equation takes into account contribution of
differential rotation into the force balance across magnetic field
lines.  The current density in the open field line zone of the
magnetosphere depends on differential rotation through condition at
the true LC, which is at $x(\psi) = c/\OmF(\psi)$ (see Sec.~2.3 in
\papI)
\begin{equation}
  \label{eq:CondAtLC}
  S \frac{d S}{d \psi} =
  2\beta \, \partial_x \psi +
  \frac{1}{\beta} \frac{d \beta}{d \psi} \left( \nabla \psi \right)^2
  \,.
\end{equation}
Changes of $\beta$ results in changes of the poloidal current density
$j$ being proportional to $dS/d\psi$.

So, the differential rotation contributes to the force balance
perpendicular to the magnetic field lines, changes the position of the
LC and affects the current density distribution, required for smooth
transition of the solution trough the LC.  Although, the differential
rotation can modify the current density distribution, it is a
quantitative question whether or not a given differential rotation
$\Omega_F(\psi)$ and the corresponding current density distribution
$j(\psi)$ could be supported by stationary cascades in the polar cap
of pulsar.  We note that $\beta(\psi)$ is a free parameter in this
problem, while $j(\psi)$ is obtained from the solution.

For solution of equation~(\ref{eq:PsrEq}) we developed an advanced
version of the multigrid code described in \papI.  Now the position of
the Light Cylinder is found self-consistently on each iteration step
and poloidal current is calculated from the equation
(\ref{eq:CondAtLC}) at the LC.  The return current flowing in the
current sheet between the closed and open field line zones is smeared
over the interval $[\PsiL,\PsiL+d\psi]$%
\footnote{$\PsiL$ is the value of the normalized magnetic flux
  function corresponding to the field line separating the closed and
  open field line domains},
and the angular velocity is continuously changing in the same interval
from some value at the last open filed line $\OmF(\PsiL)$ to $\Omega$
at $\psi=\PsiL+d\psi$.  Equation~(\ref{eq:PsrEq}) is solved in the
whole domain including the current sheet.  Smearing of the return
current and continuous changing of $\OmF$ in the current sheet allows
us to take into account the contribution of the current sheet into the
force balance across magnetic field lines at the boundary between
closed and open field line zones.

\begin{figure}
  \centering
  \includegraphics[clip,width=\columnwidth]{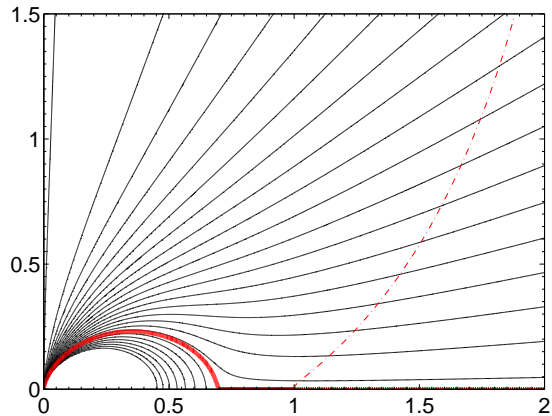}
  \caption{Structure of the magnetosphere for $\XL=0.7$ and the
    angular velocity of the open magnetic field lines
    $\OmF(\psi)=\Omega\bar{\beta}(\psi)$ (see text). Magnetic field
    lines are shown by the thin solid lines, the last closed field
    line -- by the thick solid line, the position of the Light
    Cylinder is shown by the dot-dashed line.}
\label{fig:x}
\end{figure}

\section{Main results}
\label{results}

Stationary polar cap cascades without particle inflow from the
magnetosphere would provide more or less constant current density
distribution over the polar cap \citep[see
e.g.][]{Harding/Muslimov98,Gil/Melikidze/Geppert:2003}.  However, in
configurations of the force-free magnetosphere with constant $\OmF$
the current density goes to zero near the polar cap boundaries.  Here
we try to find configurations of a differentially rotating force-free
magnetosphere with current density, which is nearly constant over the
polar cap at the NS surface.  Current density in the magnetosphere
rotating with a constant angular velocity decreases toward the polar
cap boundary because of the boundary condition $\psi(x>\XL) = \PsiL$
($x_0$ is the position of the point, where the last closed field line
intersects the equatorial plane).  In the equation for the poloidal
current (eq.~(\ref{eq:CondAtLC})) the first term on the left hand side
goes to zero for $\psi$ approaching $\PsiL$.  Hence, in order to
increase the current density near the polar cap boundaries
$d\beta/d\psi$ must be positive, i.e. the maximum value of $\beta$
should be achieved at the polar cap boundary, for $\psi=\PsiL$.
Obviously, the maximum possible value of $\beta$ is 1.  Such behavior
of $\beta$ is expected in models of the polar cap cascades too.

\begin{figure}
  \centering
  \includegraphics[clip,width=.85\columnwidth]{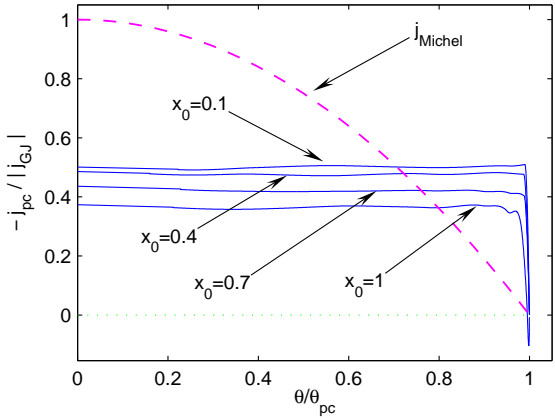}
  \caption{Current density distributions in the polar cap of pulsar
    $\PC{j}(\theta)$ corresponding to $j(\psi)\approx\bar{\jmath}$
    (see text) are plotted for different \XL{} as functions of the
    colatitude $\theta$.  $\PC{j}$ is normalized to the
    Goldreich-Julian current density $|\GJ{j}|$.  $\theta$ is
    normalized to the colatitude of the polar cap boundary \QPC.  The
    current density corresponding to the Michel's solution
    \citep{Michel73:b} is shown by the dashed line. The distribution
    for $\XL=0.1$ practically coincides with the distribution for the
    split-monopole case.}
\label{fig:jpc}       
\end{figure}

We have performed computations for different sizes of the closed field
line zone \XL.  In each of these simulations the distribution of
$\beta$ over $\psi$ was adjusted ad hoc in order to obtain nearly
constant current density over the polar cap, which approaches zero
only at field lines very close to the polar cap boundary%
\footnote{we assume that the current sheet carries only the return
  current, so $j$ changes the sign at \PsiL.}.
For any size of the closed field line zone \XL{} it is possible to
construct a set of force-free magnetospheric configurations with the
current density distribution being almost constant over the polar cap,
$j(\psi)\simeq\hat{\jmath}\equiv const$ by choosing different
$\beta(\psi)$.

One of the important properties of the obtained set of solutions is
that for each \XL{} the constant current density $\hat{\jmath}$ cannot
be made greater than some maximum value $\bar{\jmath}$.  Distribution
$j(\psi)\simeq\bar{\jmath}$ corresponds to a distribution
$\bar{\beta}(\psi)$, which achieves 1 at the polar cap boundary,
$\bar{\beta}(\PsiL)=1$.  As it was stressed above, the maximum value
of $\beta$ is achieved at $\psi=\PsiL$.  So, $\bar{\beta}$ differs
from other distributions $\beta(\psi)$ which provide constant current
density, in that it achieves the maximum possible value.  Both
$\bar{\jmath}$ and $\bar{\beta}(\psi)$ depend on \XL.

\begin{figure}
\centering
  \includegraphics[clip,width=.85\columnwidth]{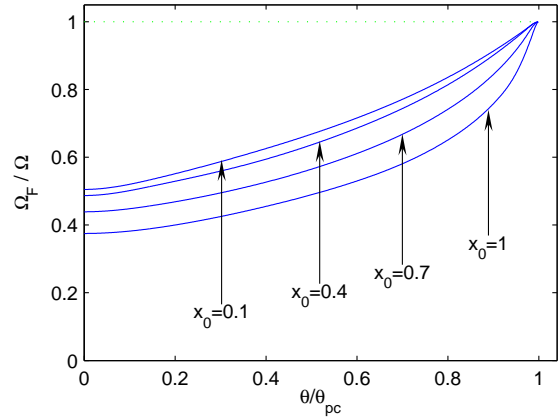}
  \caption{Angular velocities of rotation of the open field lines
    $\OmF(\theta)=\Omega\;\bar{\beta}(\theta)$ (see text) for
    different \XL{} as functions of the colatitude $\theta$ in the
    polar cap of pulsar.  $\theta$ is normalized to the colatitude of
    the polar cap boundary \QPC.  $\OmF(\theta)$ is normalized to the
    angular velocity of the NS rotation $\Omega$. The distribution for
    $\XL=0.1$ practically coincides with the distribution for the
    split-monopole case. }
\label{fig:beta}       
\end{figure}

\begin{figure}
\centering
  \includegraphics[clip,width=.85\columnwidth]{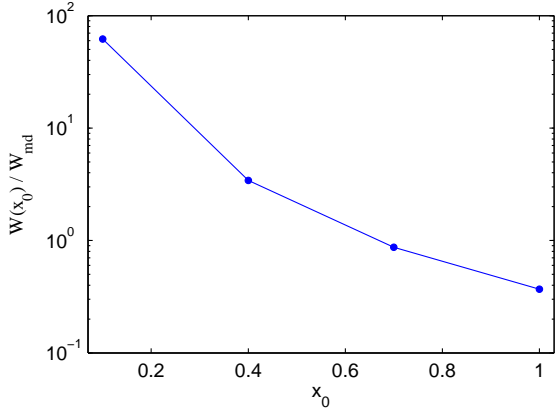}
  \caption{Energy losses of aligned rotator normalized to the
    magnetodipolar energy losses as a function of $\XL$ for
    differentially rotating magnetosphere with
    $\OmF(\psi)=\Omega\bar{\beta}(\psi)$ (see text).}
\label{fig:w}       
\end{figure}

The differential rotation in the magnetosphere of a young pulsar with
stationary polar cap cascades is rather small (see Section.~2.2 in
\papI). So, solutions with the minimal deviation of $\beta(\psi)$ from
1 would be of most interest for us.  These turned out to be solutions
with $\beta=\bar{\beta}(\psi)$, i.e. with $j\simeq\bar{\jmath}$.  One
of such magnetospheric configurations with $\XL=0.7$ is shown in
Fig.~\ref{fig:x}. Important properties of solutions with
$j\simeq\bar{\jmath}$ are:
\begin{itemize}

\item The value $\bar{\jmath}$ increases with decreasing of $x_0$,
  however it does not exceeds $\GJ{j}/2$, the half of the GJ current
  density near the NS surface (see Fig.~\ref{fig:jpc}).

\item The angular velocity of rotation of the open magnetic field
  lines has smaller deviation from $\Omega$ with decreasing of \XL,
  see Fig.~\ref{fig:beta}.  As in the case of the current density
  distribution there exist an asymptotic distribution of $\OmF(\psi)$.
  This asymptotic distribution deviates strongly from
  $\beta(\psi)\equiv{}1$.

\item The asymptotic form of $\bar{\beta}(\psi)$ and $\bar{\jmath}$
  for $x_0\rightarrow{}0$ matches the corresponding distributions in
  the split monopole configuration.

\item Energy losses of the aligned rotator increases with decreasing of
  \XL. This increase goes a bit faster than in the case of constant
  $\OmF$, cf.  Fig.~\ref{fig:w} here and Fig.~8 in \papI.

\item The total energy of the magnetosphere increases with decreasing
  of \XL.

\item In none of the considered configurations the force-free
  condition is violated in the calculation domain, i.e. the electric
  field is everywhere smaller than the magnetic field.

\end{itemize}

Let us now discuss implications of these solutions for the physics of
pulsars.

\section{Discussion}
\label{discussion}

If the polar cap cascades operate in stationary SCLF regime the
current density in the polar cap of pulsar near the NS surface is very
close to the local GJ current density. However, due to inertial frame
dragging (i.e. due to changing of the effective $\Omega$) the local GJ
charge density decreases slower with the distance than the charge
density carried by the charge-separated flow from the NS surface
\citep[see][]{Muslimov/Tsygan92}. This discrepancy achieves
$\sim{}15\%$ at a distance of several radii of the NS. So, the current
density at that distance will be by $\sim{}15\%$ less then the
corresponding local GJ current density.  Here we solve the problem in
flat space-time. The boundary conditions on the NS surface and, hence,
the physical quantities we used for normalization should be taken at
some distance from the NS, where GR effects are negligible. The GR
corrections will result in $\sim{}15\%$ higher ratio of $j$ to the
\emph{local} value of $\GJ{j}$ at the NS surface than the one obtained
in our solution.  However as it follows from our results, there is no
configurations of the magnetosphere with almost constant current
density larger than $\GJ{j}/2$.  Hence, stationary force-free
magnetosphere of an aligned rotator can not be supported by stationary
polar cap cascades operating in SCLF regime.

Polar cap cascades operating in \citet{Ruderman/Sutherland75} regime
with almost vacuum electric field in the gap allow larger deviations
of the current density from $\GJ{j}$ as well as larger deviation of
$\beta$ from 1.  For older pulsar, when $\beta$ could be rather small,
even configuration with $x_0$ approaching 1 could be supported by the
polar cap cascades.  In this case the characteristic current density
flowing through the gap must be also much less than the GJ current
density.

For the partially screened polar gap (PSG)
\citep{Gil/Melikidze/Geppert:2003} the presence of a strong multipolar
component of the magnetic field in the polar cap of pulsar is
essential.  In this model the current density at the NS surface for
young pulsars should be close to the local value of $\GJ{j}$.
However, the local GJ charge density is determined by the non-dipolar
magnetic field.  For this field the angle between $\vec{B}$ and
$\vec{\Omega}$ can be rather large and the GJ charge density will be
accordingly less.  At some distance, where magnetic field becomes
dipolar and the angle between $\vec{B}$ and $\vec{\Omega}$ is small,
the current density of the flow would become smaller than the local GJ
current density.  This effect can be much stronger than the difference
between $j$ and $\GJ{j}$ due to inertial frame dragging. Hence, for
some specific configurations of the local magnetic field at the NS
surface, stationary polar cap cascades could support a stationary
force-free configuration of the magnetosphere.

So, aligned pulsars with polar cap cascades operating in SCLF regime,
young pulsars with ``Ruderman-Sutherland'' cascades as well as pulsars
with PSG and arbitrary surface magnetic field configuration can not
have both polar cap cascades operating in stationary regime and
a stationary force-free magnetosphere. From energetic point of view,
configurations with non-stationary polar cap cascades would seem to be
preferable.  The energy of the magnetosphere increases with decreasing
of \XL. The magnetosphere would try to achieve the configuration with
the smallest possible energy, i.e. with the largest possible \XL,
compatible with the physical conditions set by the polar cap cascades.
If there are several possible self-consistent configuration of the
magnetosphere and polar cap acceleration zone, the configuration with
the largest \XL will be realized.  With aging of the pulsar the
conditions in the cascades change.  This would result in changing of
the magnetospheric configuration and, hence, in changes of energy
losses relative to the corresponding magnetodipolar losses.  This
yields the breaking index different from 3. However, more detailed
study of polar cap cascade properties is necessary in order to
construct self-consistent magnetospheric configurations and to obtain
the value of the breaking indexes.  The current adjustment is
necessary also for inclined rotator and in this case similar effects
would be present too.

More detailed description of the results is given in
\citet{Timokhin06_psreq2}.

\begin{acknowledgements}
  This work was partially supported by RFBR grant 04-02-16720, and by
  the grants N.Sh.-5218.2006.2 and RNP.2.1.1.5940
\end{acknowledgements}

\bibliography{psreq}   

\begin{thebibliography}{10}

\bibitem[\protect\citeauthoryear{Beskin \& Malyshkin}{Beskin \&
    Malyshkin}{1998}]{Beskin/Malyshkin98} {Beskin}, V.S., {Malyshkin},
  L.M.  \mnras \textbf{298}, 847 (1998)

\bibitem[\protect\citeauthoryear{Bucciantini et al.}{2006}]{Bucciantini06}
  {Bucciantini}, N., {Thompson}, T.A., {Arons}, J. et al.
  \mnras \textbf{368}, 1717 (2006)


\bibitem[\protect\citeauthoryear{Contopoulos}{2005}]{Contopoulos05}
  {Contopoulos}, I. 
  \aap \textbf{442}, 579 (2005)


\bibitem[\protect\citeauthoryear{Contopoulos et. al}{1999}]{CKF}
  {Contopoulos}, I., {Kazanas}, D., {Fendt}, C. 
  \apj \textbf{511}, 351 (1999)

\bibitem[\protect\citeauthoryear{Gil et al.}{2003}]{Gil/Melikidze/Geppert:2003}
  {Gil}, J., {Melikidze}, G.I., {Geppert}, U.
  \aap \textbf{407}, 315 (2003).

\bibitem[\protect\citeauthoryear{Goldreich \& Julian}{1969}]{GJ}
  {Goldreich}, P., {Julian}, W.H.
  \apj \textbf{157}, 869 (1969)

\bibitem[\protect\citeauthoryear{Goodwin et al.}{2004}]{Goodwin/04}
  {Goodwin}, S.P., {Mestel}, J., {Mestel}, L., {Wright}, G.A.E.
  \mnras \textbf{349}, 213 (2004)

\bibitem[\protect\citeauthoryear{Gruzinov}{2005}]{Gruzinov:PSR}
  Gruzinov, A.
  Phys.Rev.Lett. \textbf{94}, 021,101 (2005)

\bibitem[\protect\citeauthoryear{Harding \& Muslimov}{1998}]{Harding/Muslimov98}
  {Harding}, A.K., {Muslimov}, A.G.
  \apj \textbf{508}, 328 (1998).

\bibitem[\protect\citeauthoryear{Komissarov}{2006}]{Komissarov06}
  {Komissarov}, S.S.
  \mnras \textbf{367}, 19 (2006).

\bibitem[\protect\citeauthoryear{McKinney}{2006}]{McKinney:NS:06}
  {McKinney}, J.C.
  \mnras \textbf{368}, L30 (2006)

\bibitem[\protect\citeauthoryear{Michel}{1973}]{Michel73:b}
  {Michel}, F.~C.
  \apj  \textbf{180}, 207 (1973)

\bibitem[\protect\citeauthoryear{Muslimov \& Tsygan}{1992}]{Muslimov/Tsygan92}
  {Muslimov}, A.G., {Tsygan}, A.I.
  \mnras \textbf{255}, 61 (1992)

\bibitem[\protect\citeauthoryear{Okamoto}{1974}]{Okamoto74}
  {Okamoto}, I. 
  \mnras \textbf{167}, 457 (1974)

\bibitem[\protect\citeauthoryear{Ruderman \& Sutherland}{1975}]{Ruderman/Sutherland75} 
  {Ruderman}, M.A., {Sutherland}, P.G.  
  \apj \textbf{196}, 51 (1975)

\bibitem[\protect\citeauthoryear{Spitkovsky}{2006}]{Spitkovsky:incl:06}
  {Spitkovsky}, A.
  \apjl \textbf{648}, L51  (2006)

\bibitem[\protect\citeauthoryear{Timokhin}{2006}]{Timokhin06_psreq2}
  Timokhin, A.N.  
  \newblock in preparation (2006)

\bibitem[\protect\citeauthoryear{Timokhin}{2006}]{Timokhin2006:MNRAS1}
  {Timokhin}, A.N.
  \newblock \mnras \textbf{368}, 1055--1072 (2006).

\end{thebibliography}

\end{document}